\documentstyle[12pt,aaspp4]{article}
\def \kms       {km~s$^{-1}~$}
\def \kmsMpc    {km~s$^{-1}$~Mpc$^{-1}~$}

\begin{document}

\title{Progressive Star Bursts and High Velocities in the
Infrared Luminous, Colliding Galaxy Arp 118}

\author{Susan A. Lamb\altaffilmark{1}, Nathan C.
Hearn\altaffilmark{2},
and Yu Gao\altaffilmark{3}}
\affil{University of Illinois at Urbana-Champaign}
\authoraddr{Department of Physics, Loomis Laboratory of Physics,
1110 W. Green Street, Urbana, IL 61801, USA}
\authoremail{slamb@astro.uiuc.edu, n-hearn@astro.uiuc.edu,
gao@astro.uiuc.edu}

\altaffiltext{1}{Department of Physics and Department
of Astronomy, UIUC}
\altaffiltext{2}{Department of Physics, UIUC}
\altaffiltext{3}{Laboratory for Astronomical Imaging,
Department of Astronomy, UIUC}


\begin{abstract}

In this paper we demonstrate for the first time the connection
between the spatial and temporal progression of star formation
and the changing locations of the very dense regions in the gas 
of a massive disk galaxy (NGC~1144) in the aftermath of its 
collision with a massive elliptical (NGC~1143). These two galaxies
form the combined object Arp 118, a collisional ring galaxy system.
The results of 3D, time-dependent, numerical simulations of the 
behavior of the gas, stars, and dark matter of a disk galaxy and 
the stars and dark matter in an elliptical during a collision are
compared with multiwavelength observations of Arp 118. The 
collision that took place approximately 22 Myr ago generated a 
strong, non-linear density wave in the stars and gas in the disk
of NGC~1144, causing the gas to became clumped on a large scale. 
This wave produced a series of superstarclusters along arcs and 
rings that emanate from the central point of impact in the disk. 
The locations of these star forming regions match those of the 
regions of increased gas density predicted the time sequence of 
models.

The models also predict the large velocity gradients observed 
across the disk of NGC 1144. These are due to the rapid radial 
outflow of gas coupled to large azimuthal velocities in the 
expanding ring, caused by the impact of the massive intruder.

\end{abstract}

\keywords{galaxies: individual (Arp 118) --- galaxies: interactions 
 --- galaxies: kinematics and dynamics --- galaxies: starburst ---
galaxies: star clusters --- hydrodynamics}

\clearpage

\section{Introduction}

Optically-identified, collision-produced 'ring galaxies' usually
display vigorous star formation in an expanding ring or arc of high
density gas (e.g., Joy \& Harvey, 1987; Joy \& Ghigo, 1988;
Appleton \& Marston, 1997; see Appleton \& Struck-Marcell, 1996). 
This recent star formation often dominates the appearance
of these galaxies, with H$\alpha$ emission tracing the giant HII regions 
(e.g., Hippelein, 1989a) and outshining the nucleus unless the latter is
active. Near-infrared images of these systems show that the older stellar 
disk population has also been swept into an outwardly expanding, strong
density wave.

The density disturbance produced by a collision through (and roughly
perpendicular to) a galaxy's disk propagates outward through a rotating
disk of gas and stars, which is itself at first contracting and then
expanding as the collision proceeds. This superposition of material 
motions and wave propagation produces a pattern of both closed loops 
and open-ended arcs of relatively high density gas. Shocks can occur 
in these regions because of the high relative velocities that are
produced in the flows. The higher density features are well delineated
in observations of these systems and in the combined N-body/hydrodynamic
models that we and others have produced. The models can be exploited to
further our understanding of the star formation that is triggered by a 
galaxy collision, a process that has likely occurred in many galaxies
over their lifetimes and may have been very frequent earlier in the 
universe (see Lavery {\it et al.}, 1996).

Here we compare our numerical models of these types of galaxy systems
with observational data on Arp 118, one particular IR-luminous, 
gas-rich example of a collisionally produced ring galaxy. This system 
consists of a strongly distorted disk galaxy and an elliptical in 
close proximity. Hippelein (1989a) had difficulty in explaining the 
extreme velocity gradient and complex morphology in the Arp 118 
system using a his simple picture of a collision between a gas-rich 
spiral and an elliptical. However, Gao {\it et al.} (1997) have 
observed that Arp 118 contains a large amount of molecular gas 
distributed exclusively along the ring formations (the first 
high resolution CO observations ever made of a ring galaxy) , and 
that the velocity structure in this gas is kinematically consistent 
with the simple collisional model.

In this paper, we show that, using fully dynamical, 3D models, we 
can reproduce the morphology of the disturbed disk galaxy in the 
pair and the approximate relative positions of the two 
galaxies. The 'best-fit' model for the Arp 118 system was chosen 
from a grid of simulations produced by Gerber, Lamb, \& Balsara 
(1996). These simulations explore the results of face-on 
collisions (collisions parallel to the spin-axis of the disk)
between an elliptical galaxy and a disk galaxy. This dynamical 
treatment confirms the correspondence between the models and the 
observed velocity structure. We use the chosen simulation to
constrain the timescales for star formation in Arp 118 and to
explore the history of the collision by comparing the predicted 
results of such a collision with current observations.

\section{Computational Method and Galaxy Models}

The numerical experiments are comprised of combined 
N-body/hydrodynamics, 3D numerical simulations of collisions 
between a rotating disk galaxy (composed of gas, stars, and 
dark matter) with a spherical galaxy composed of stars and 
dark matter only. Disk galaxy to elliptical mass ratios of 1:1, 
4:1, and 10:1 have been calculated with impact parameters 
ranging from zero (head-on) to slightly larger than the disk radius.
The gas dynamics is treated using the smoothed particle hydrodynamics 
(SPH) method, and the gravitational forces are calculated using 
standard particle-mesh (PM) techniques. Both the stellar and SPH 
particles contribute to the gravitational potential which is 
calculated on a cubic grid with 64 points along each side. Further
details of the code can be found in Balsara (1990), where tests of 
it are presented. All computations were performed on the Cray-2 
supercomputer at the National Center for Supercomputing 
Applications (NCSA) at the University of Illinois at Urbana-Champaign.

The spherical, gasless elliptical galaxy model can be thought of as 
type E0, and the disk galaxy model can be thought of as an Sc galaxy. 
The spherical components of the disk galaxy (the halo stars and dark 
matter) and the elliptical were represented by King models. In these 
simulations the exponential disk, which is comprised of both stars 
and gas, has a mass two-fifths that of the halo and is cut off in 
radius at 4.4 scale lengths. The mass of the gas is one tenth of the 
total mass of the disk. The gas is modeled using approximately 22,000 
SPH particles which are initially placed in circular orbits around the
center of the disk with the gas density distribution set proportional
to that of the disk stars. The disk stars and the halo of the disk 
galaxy are each represented by 25,000 (N-body) particles, whereas the 
elliptical is represented by 10,000 particles.

The gas is represented by a single phase which is used to model the 
large-scale flows of the gas and to allow a determination of
where and when the high densities and shocks occur on a
scale larger than that of individual clouds in the interstellar
medium. The gas density in the central plane of the galaxy
is such that 0.025 amu cm$^{-3} < \rho_{\rm gas}<$ 2 amu cm$^{-3}$,
and the temperature is in the range 8000 K $<~T/\mu~<~6\times10^5$ K. 
It is assumed that the time scale for radiation processes that cool 
the gas behind shock fronts is shorter than the calculational time 
step, that is, that the gas is isothermal. Full details 
of the computational method and the starting models can be found in
Gerber (1993) and Gerber, Lamb, \& Balsara (1996).

In general, the simulations give high volume densities in the gas at
those places where kinetic theory would predict orbit crossing.
If we assume that star formation in colliding galaxies is
triggered by shocks in the gas in high density regions, 
then we can predict the locations of intense bursts of star formation.
The volume density of the gas can increase by a factor of
about 100 in the ring, while the line-of-sight surface density
increases in these regions are much more modest. (The increase in 
surface density of the stars in the ring is only a factor of a few, 
leading to a much less pronounced ring formation.) Thus we do not 
necessarily expect a high correlation between the locations of star 
bursts in a real galaxy with the locations of high gas {\it surface} 
density in the models (nor of that in the systems themselves). However, 
we have previously predicted and demonstrated that there is a
good correlation between the regions of high {\it volume} density and
shocks in the models and the locations of clumps of star formation
in a colliding system (see the study of the Arp 147 system by
Gerber, Lamb, \& Balsara, 1992).

\section{Observations and Model Fits}

The Arp 118 interacting system is comprised of two galaxies:
an elliptical (NGC~1143) and a disturbed disk galaxy (NGC 1144)
which contains a very extended starburst ring and connected arc.
The orientation of the two galaxies and the projected separation 
of their nuclei are
shown in Fig. 1 (plate 1), which displays a Hubble Space Telescope
image of the pair taken with the WFPC 2 camera through
filter F606W (V-band) by Malkan {\it et al.} on Nov 22,
1995\footnote{Based on observations made
with the NASA/ESA Hubble Space Telescope, obtained
from the data archive at the Space Telescope Science
Institute. STScI is operated by the Association of
Universities for Research in Astronomy, Inc. under the
NASA contract  NAS 5-26555.}.
These two galaxies have an angular separation of 40$"$, which
is equivalent to a separation of 22 kpc on the sky (assuming
H$_{0}=75$ \kmsMpc throughout), and a relative velocity between
their nuclei of 300 \kms along the line of sight.
NGC 1144 is a luminous
infrared galaxy with a total luminosity of approximately
$3\times10^{11}~$L$_{\odot}$ and L$_{IR}=2.5\times10^{11}~$L$_{\odot}$
(Appleton \& Struck-Marcel, 1987), a very
strong CO(1-0) emission (L$_{CO}=10^{10}~$L$_{\odot}$,
Gao {\it et al.}, 1997), and an extreme velocity
gradient across the ring of 1100 \kms (see Hippelein, 1989a, b, 
and Fig. 2a which is reproduced here from these papers).
The figure does not denote the observed velocity
for the nucleus of NGC 1144. Rather, the velocity contours on
either side of the nucleus indicate a large
velocity gradient across the nucleus. The nucleus is displaced 
from the center of the ring structure and is classified as 
Seyfert 2 (Hippelein, 1989a; Osterbrock \& Martel, 1993).

These unusual properties have led some to suggest that NGC 1144
is itself the result of a merger between two galaxies (Hippelein,
1989a; McCain \& Freeman, 1994; Gao, 1996). Here we show that we 
can explain the disturbed morphology, the extreme velocity 
gradients, and the morphological relationships between the 
emitting regions (radio, H$\alpha$, and CO) as resulting from 
a collision between a rotating disk galaxy and a somewhat 
less massive elliptical which passed through the disk 
approximately 22 Myr ago at a slight angle to the perpendicular 
to the disk. The detailed modeling does not support the merger
suggestion.

From the form of the morphological disturbances in this galaxy,
we estimate from the model fit that the impact took place at a 
distance of 7.7 kpc
from the center of the disk, where the disk is approximately
12 kpc in radius. However the impact point would appear a little
closer to the nucleus in the projected view of the model galaxy
due to the tilt at which we view the real NGC 1144.
The actual observed distance between the nucleus and the
supposed impact point in NGC 1144 is close to 4 kpc. We suspect
that this proximity is only apparent, and is due to
a projection effect of the nucleus-ring
separation onto the plane of the sky, where the nucleus
has been displaced by the collision out of the original plane of
the disk. Such projection effects are
moderately common in collisionally produced ring
galaxies, yielding apparently very off-center nuclei
or nuclei that appear buried in a ring (Gerber,
Lamb, \& Balsara, 1992; Lamb {\it et al.}, 1998).

The overall dynamics and morphology of the Arp 118 system are
well represented by a penultimate model in one of our N-body/SPH
simulations of an off-center collision in which the disk galaxy
has a mass four times that of the intruding elliptical.
In particular, the most notable large-scale feature of the system,
the extensive arm of compressed gas edged with a dark dust lane
(as seen in the WFPC2 image), is very well matched by our model.
In order to fit the observations, the model galaxies
were viewed at an angle of 47 degrees from the perpendicular
to the plane of the disk, with the intruding elliptical
located between us and the disk galaxy. (Thus a significant 
relative velocity across the sky, as well as in the line-of-sight, 
is expected for these two galaxies). Using this projection angle,
we estimate the distance between the two galaxies
to be 35 kpc, and their relative space velocity to be
around 400 \kms. This velocity is almost identical to that of 
the two model galaxies at the end of our matching simulation.

The relative R-band intensities of NGC 1144 and NGC~1143
(Hippelein, 1989a) as well as the near-IR flux densities,
mainly the K band (Joy \& Ghigo, 1988) suggest
that the mass ratio of these two galaxies is approximately 2:1,
but the true ratio is somewhat uncertain due to lack of
information on the extent of the dark halos. However our sets of
simulations provide strong evidence that the mass ratio is not 1:1, as
these simulations develop morphology that is distinctly different from
that observed in NGC 1144.

The mass ratio of the two galaxies involved in a collision,
together with the mass of the effected galaxy
determine the physical timescale for the propagation of the
density wave through the disk. Consequently, the
distance between star formation regions will depend on
these masses. The new stars in these systems tend to be
formed in large (linear dimension of approximately 1 kpc)
clumps and thus a high speed wave will allow the star
formation regions to be spread in space, that is, to be
distinct from each other rather than overlap. Thus a massive
gaseous disk hit by a massive intruder provides the best
opportunity to detect and study the star formation over
the collision history. Further, it is possible to put a
rough constraint on the mass ratio of the two galaxies
through a careful comparison of the sequence
of models in a simulation with multiwavelength maps of the
system that provide a relationship between the locations and
times of star formation. When we make these comparisons for
the Arp 118 system we conclude that the disk galaxy is more 
massive than the elliptical by a factor of two to four.

The physical mass and size of the disk galaxy is used
to determine the relevant physical interpretation of the
timescale in the simulation. Gao {\it et al.} (1997) have estimated
the mass within the ring/arc of NGC~1144 using dynamical
arguments and an average CO(1-0) emission ring radius of
6 kpc to be approximately $2\times10^{11}~$M$_{\odot}$ (this length
and mass are dependent on H$_{0}$), which is about four times the
mass of the disk of the Milky Way galaxy.  However when making a 
comparison with our galaxy simulations it must be noted that
the total simulated galaxy mass includes contributions from both
a disk of gas and stars and a halo of stars and dark matter, 
with relative masses of 2:5. Thus we wish to compare the observed 
properties of NGC~1144 with a simulation in which the mass of 
the disk galaxy is $5\times10^{11}~$M$_{\odot}$.

With this scaling of the mass and the observed diameter of
the outer CO feature and H$\alpha$ ring ($\sim 20$ kpc), we
find that the velocity gradient across
our model viewed from the relevant angle is 950 \kms
(see Figure 2b). This compares favorably with the velocity
gradient of 1100 \kms found by Hippelein (1989a, b)
in H$\alpha$ (see Figure 2a) and by Gao {\it et al.} (1997) in CO(1-0).
The difference in velocity is very likely due to the real
intruder having a higher mass relative to the disk galaxy
than that in our simulation. A higher mass impact produces
larger radial outflows (see Gerber, Lamb, \& Balsara, 1996)
that could easily contribute an extra 150 \kms velocity
difference across this system. 
The numerical velocity contours have twice the resolution of
the observational ones displayed here, and extend to a larger
radius in the disk.  Consequently, the features at $+500$ and
$-400$ \kms in the model are not well-matched by these
observations.  The more recently determined, higher resolution
velocity contours of McCain (1998) show a much closer fit to our
model contours.

\section{Multi-Wavelength Morphology and the Star Formation Time Sequence}

We determine the sequencing of the various star formation
episodes in NGC~1144 by matching the models at three different
stages of the simulation with observations in three different 
wavelength bands, each of which corresponds to a well-defined epoch of
star formation. The radio emission and H$\alpha$ measurements
of a starbursting galaxy are an indication of the star formation activity
that occurred at earlier times, the H$\alpha$ being from the more
recently formed stars. The CO observations provide information about
those regions of the gas that are currently dense and likely to
form stars shortly. In making these comparisons a first
step is to fix the timescale appropriate for a correct
interpretation of the numerical simulations. The mass and linear 
size of NGC~1144 set the computational time unit in our simulation 
at 0.56 Myr. Thus a comparison of the information in the models 
with the morphology and, where available, kinematics in various 
wavelength bands allows the timing of the various star formation 
episodes to be deduced.

In Figures 3b, 4b, and 4c we depict the gas density at different
epochs in a manner that emphasizes the volume density
of the gas rather than the surface density. We bin SPH
particles on a spatial scale of 1.1 times the SPH smoothing
length used in the original computations, and then use the
resulting mass per unit box volume to obtain a local density.
We then plot only those volume densities that are above a 
specific minimum, chosen to display the features of interest.
Here we choose to display densities larger 
than $9.4 \times 10^{6}$ M$_\odot$ kpc$^{-3}$ to show increases for
the outer parts of the disk.  These density isosurfaces provide a 
more accurate sense of
the regions in which the star formation is expected to take place 
than would a plot of surface mass density, because the collision can 
cause the gas disk to be spread considerably in the direction 
perpendicular to the disk and be warped out of the original 
disk plane.

The time since closest approach for this galaxy pair is
sufficiently long for there to have been opportunity for the
approximately 15 M$_{\odot}$ and greater mass stars formed shortly
after the collision to have reached the supernova stage (Lamb,
Iben, \& Howard, 1976). The off-nucleus synchrotron radio emission
observed by Condon {\it et al.} (1990; also Joy \& Ghigo, 1988, and
references therein) is evidence of these supernovae. Figures 3a and
3b (plate 1) compare the morphology of the radio emission
with a model from the simulation that corresponds to a time 3.4 Myr
after closest approach and 19 Myr ago. Note the fairly-compact,
dense regions disconnected from, but close to, the nucleus
in both figures. At the time when the relevant
massive stars were formed, the intruding nucleus had
caused a strong density increase in a volume centered on the region
in the disk through which the intruder's center of mass passed.
The swift inflow of gas into this region would also have
produced strong shocks. The first collision-produced star 
formation region has been spread azimuthaly by the 
subsequent differential rotation of the disk. 

The H$\alpha$ emission morphology is best matched
by a model that corresponds to a time 16 Myr after closest
approach and 7 Myr ago (see Figure 4b). Both the H$\alpha$ observations
and the model display a fairly-complete ring of large radius encircling
the nuclear region and spread considerably in the northwest direction.
Note the large buildup of gas in the southeast section of the model 
which matches that found in observations, as well as the thin part 
of the ring clearly
detached from the nucleus in the southwest part of the model. Again, 
we suggest that the locations of the H$\alpha$ producing stars are not 
now completely coincident with the current regions of highest gas density 
because the density pattern in these stars reflects the density
pattern in the gas at a slightly earlier epoch.
A similar displacement between the H$\alpha$
producing stars and the gas has recently been reported for the Antennae
galaxies by Mirabel {\it et al.} (1998).

The CO(1-0) emission comes from present dense gas and should
trace the current line-of-sight high surface density regions. These
correspond well to the high volume density regions displayed in 
the model. The best fit to the observations of Gao {\it et al.} 
(1997) is with a model that occurs towards the end of our 
simulation (Figure 4c), corresponding to 22 Myr after closest approach.
The CO contours (Gao {\it et al.} 1997) are shown in Fig. 4a (plate 2)
overlayed on the H$\alpha$ image of Hippelein (1989a). The model chosen
exhibits three key features found in the CO observations: an incomplete
ring of dense gas with a gap in the west and southwest;
large regions of dense CO gas in the south and southeast (including
a region which extends quite far to the south); and a spray of more 
diffuse gas to the northwest. The lower density gas is not displayed 
in this figure. It fills in those regions not containing the dense, 
CO-emitting gas and we anticipate that it is detectable in H I.

\section{Conclusions}

The simulation presented here is comprised of a sequence of models
that follows the evolution of a pair of galaxies through and after
a collision. One of the later models closely matches the observed 
morphology of the CO component in Arp 118 and its velocity field. 
Earlier models in the sequence provide a good match with the present 
radio continuum and H$\alpha$ emission, indicating that the sequence of
post-collision star formation can be traced and timed by comparison
to simulations of encounters between two massive galaxies with masses 
similar to those observed.

Stars, particularly massive stars, are formed in the cores of
giant molecular clouds in the highest density regions. Both
multiwavelength observations and our numerical models of Arp 118
indicate that strong shocks in the gas together with large increases 
in the gas volume density are associated with star formation over
volumes of 1 kpc$^3$. The observed morphology of the regions of dense 
gas and the clustering observed in the stars emitting H$\alpha$, 
which were formed in the gas since the collision, suggest that the 
observed clumping of the young stars results from a clumping of the 
densest gas on the same scale. The recent work by Marston \& Appleton 
(1995) and Appleton \& Marston (1997) also provides evidence that 
the clumping observed in the optical images of collisionally produced
 ring galaxies is not due to patchy dust obscuration, because the same 
clustering is also observed in the near infrared. Gas clumping on 
this same scale is found in the numerical simulations, suggesting that 
there is a global explanation for the observed morphology of the dense 
gas and the resulting giant stellar formations in these systems. The 
simulations show a relatively small perturbation (clumping) in the 
density of stars. We therefore predict that the distribution of the 
old stars in these systems will be observed to be smoother than that 
of the gas, although the models show that the stars are driven into 
a wide ring, and sometimes even a second, inner ring, by the collision 
(see Lamb {\it et al.} 1998).

This study of the Arp 118 system demonstrates that a careful comparison
between high resolution observations and detailed models can yield  
insight into the sequence of star formation that takes place 
in a gas-rich galaxy after a major collision. The intensity and 
location of the starburst at any particular epoch will depend upon the 
speed with which density waves are propagating through the expanding 
disk. Such quantities can now be predicted quite accurately from 
current models of colliding galaxies. Thus global star formation (on 
the scale of several hundred parsecs) as it occurs in these systems 
at the current epoch can be investigated more thoroughly than previously. 
The rate of galaxy collisions in the past was larger than it is today, 
due to the greater overall density, thus a considerable portion of the star
formation that took place in young disk galaxies at earlier epochs was 
likely triggered by galaxy collisions. We expect, therefore, that studies 
like the one reported here will help in understanding this earlier star 
formation and its current consequences.

\acknowledgements
We thank Laird Thompson for fruitful discussions and for comments
on the manuscript. We also thank J. Condon for providing digital 
radio images of Arp 118. SAL acknowledges the support of the 
University of Illinois Research Board and the National Center 
for Supercomputer Applications, which is part of the National 
Computational Science Alliance (NCSA). SAL and NCH also acknowledge 
use of the Renaissance Education Laboratory and the Numerical 
Laboratory at NCSA. YG's research at the Laboratory for Astronomical 
Imaging in the Astronomy Department at the University of Illinois 
is funded by NSF grants AST93-20239 and AST96-13999, and by the 
University of Illinois.

\clearpage

\begin{center} {\large \bf Figure Captions}
\end{center}

\begin{figure}[hp]

\caption[junk]{(Plate 1) Hubble Space Telescope
image of Arp~118 taken with the WFPC 2 camera showing NGC~1144
on the left and NGC~1143 on the right. The distance to the
system is 118 Mpc and the projected separation
between the two galaxies is 22 kpc (H$_{0}$ = 75 \kmsMpc).}

\end{figure}

\begin{figure}[hp]

\caption[junk]{(a.) Observed velocity contours in H$\alpha$
(Hippelein,1989a, b). Note the large gradient in nuclear region 
and the extended low velocity region above and to the right 
(Northwest) of the nucleus (position denoted by $\times$).

(b.) Simulation gas (SPH particle) mean line-of-sight velocity
at 22 Myr after collision. Line-of-sight velocities
from the simulation, given in \kms, closely resemble
those found by Hippelein. (Most positive velocity, away 
from observer, is to left of figure.
The most negative, towards observer, is on the right, with the
position of the nucleus again denoted by an $\times$.)}

\end{figure}

\begin{figure}[hp]

\caption[junk]{(Plate 1) (a.) Radio continuum map (Condon 
{\it et al.}, 90). Note the presence of a strong radio 
source above and to the left of the nucleus (but still 
close to the nuclear region). We interpret this region 
as indicating the impact point in the disk. 

(b.) Gas (SPH particle) volume density in the simulation at 
3.4 Myr after the collision. A localized increase 
in gas mass density to the left of the nucleus (about half-way 
between the nucleus and the edge of the disk) indicates
the point of impact in the model. The co-ordinates along the axes 
show computational length units for the simulation, where one unit
scales to 2.2 kpc for the Arp 118 system.}

\end{figure}

\begin{figure}[hp]
\caption[junk]{(Plate 2) (a.) CO(1-0) emission map contours 
(Gao {\it et al.}, 1997) superimposed on H$\alpha$ image 
of Arp 118 (Hippelein, 1989a).
The large ring of cold gas is broken (no significant CO emission)
in the lower-right section and there are extended regions of strong
signal below and to the left of the nucleus. However, the H$\alpha$
image shows an extended complete ring of material present around the
nucleus with more complicated features in the near-nuclear region.
We note the considerable clumping in the H$\alpha$-producing 
stellar density and the CO gas.

(b.) Gas (SPH particle) volume density in the simulation at a post 
collision time of 16 Myr. The pronounced ring indicates those 
regions that are most likely to have experienced star formation 
at this past epoch and match the current H$\alpha$ image well.

(c.) Gas (SPH particle) volume density in the simulation at a post 
collision time of 22 Myr. The large ring of gas in the model 
is also broken in the lower-right section.
A dense region of gas exists to the left of the nucleus,
accompanied by a spray of moderately-dense material south of
(below) the nucleus.  A strong tail of lower-density gas has
formed far from the nucleus towards the Northwest (upper-right).
The co-ordinates along the axes in b. and c. show computational
length units for the simulation, where one
unit scales to 2.2 kpc for the Arp 118 system.}

\end{figure}

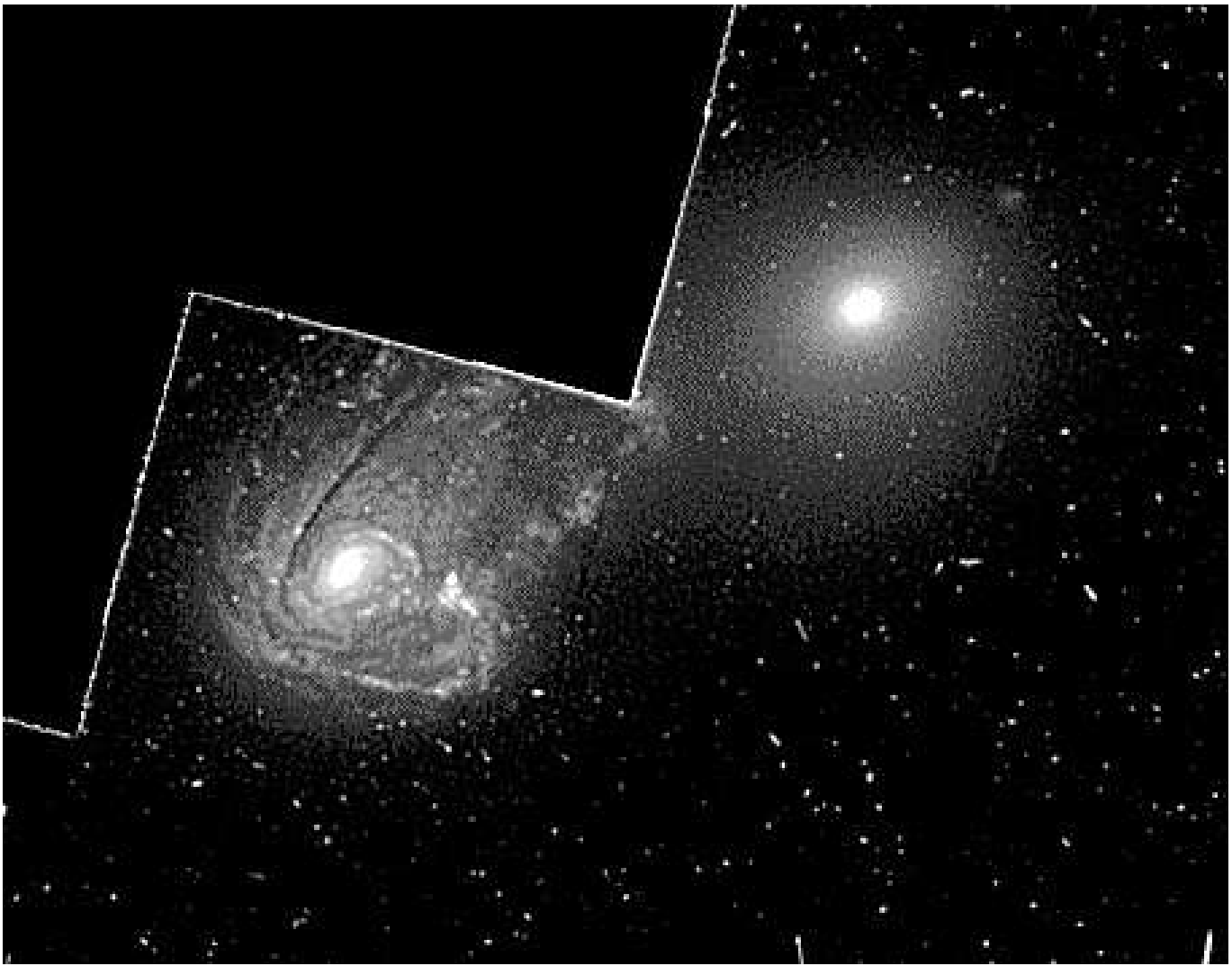
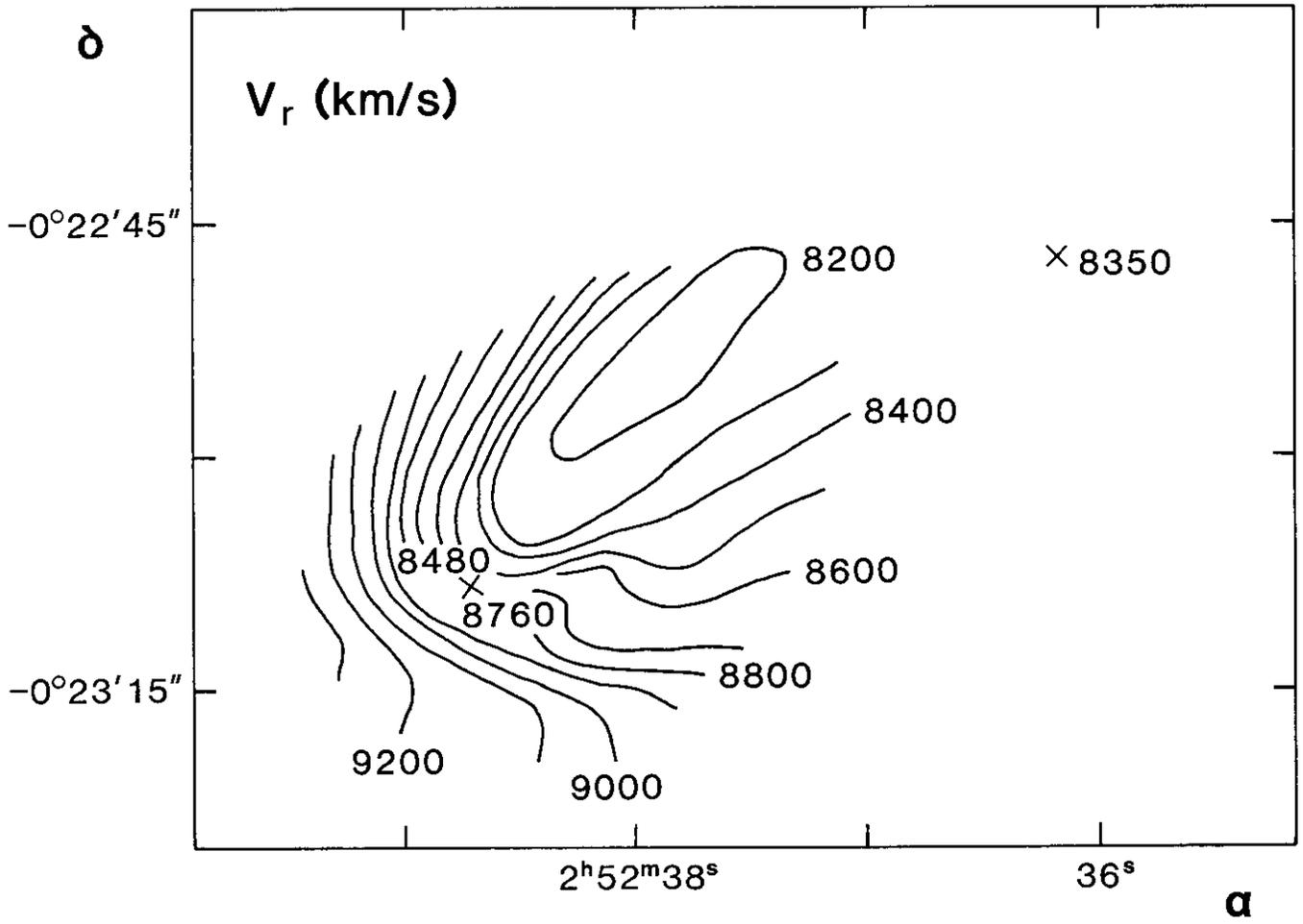
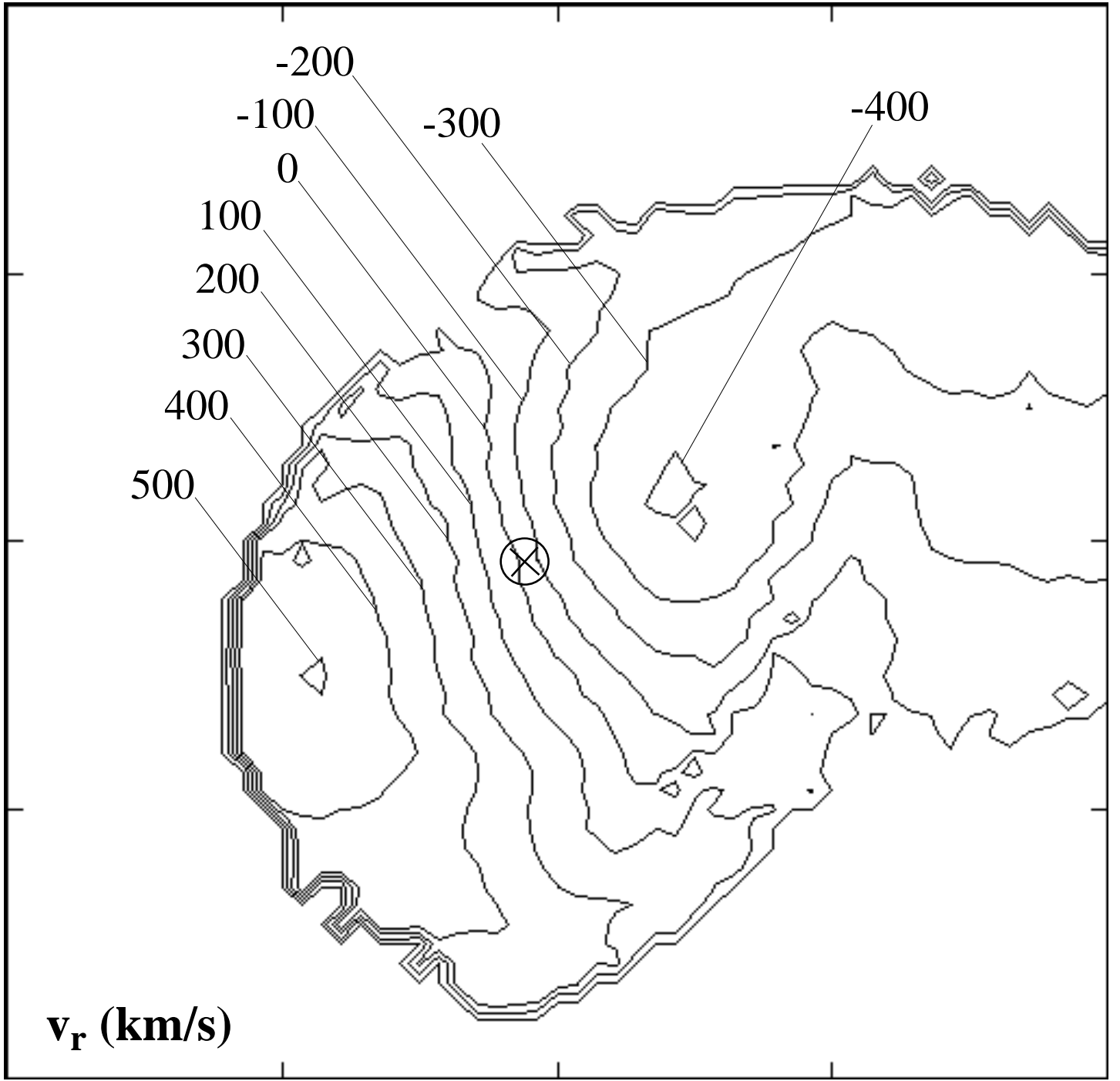
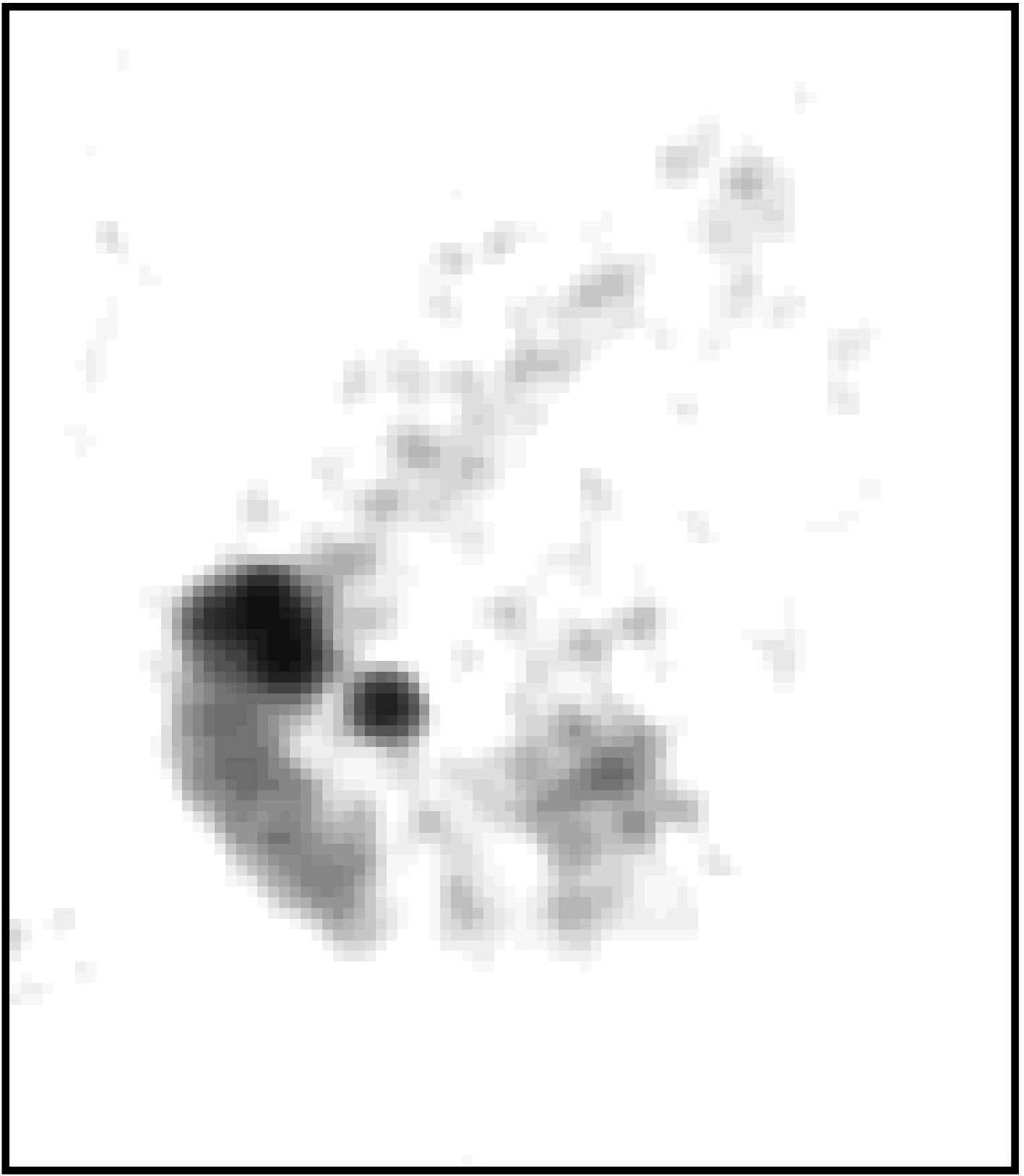
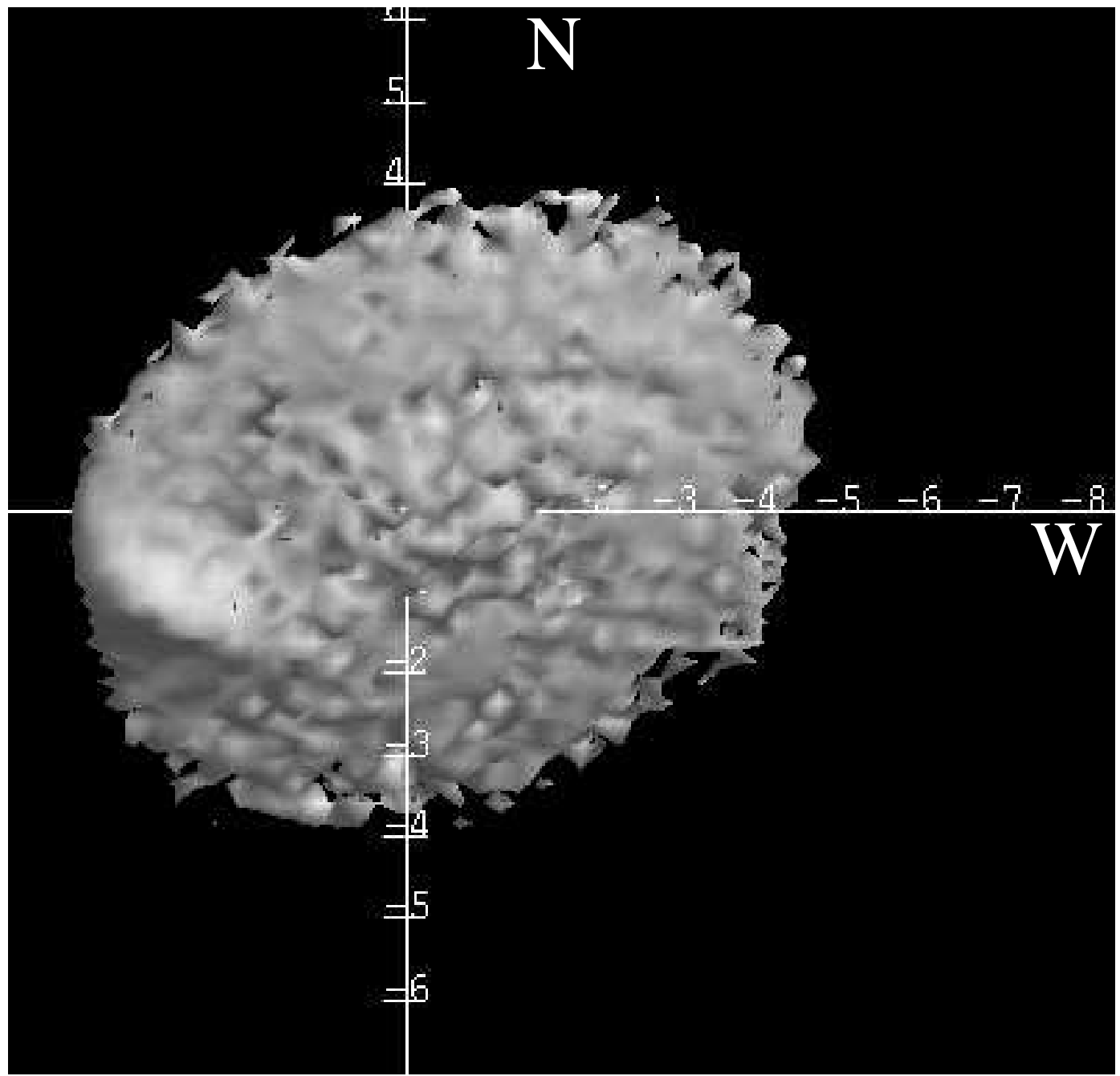
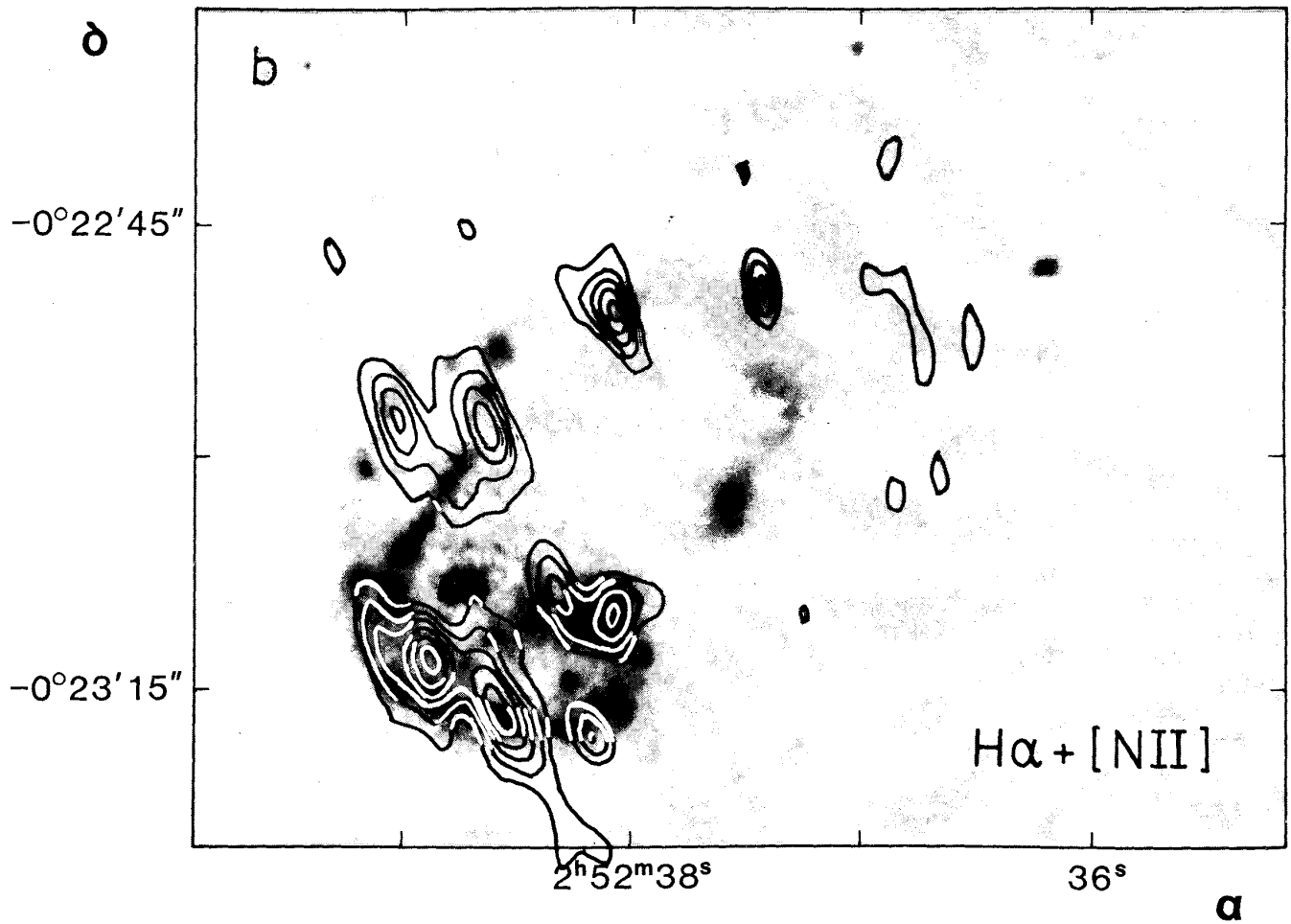
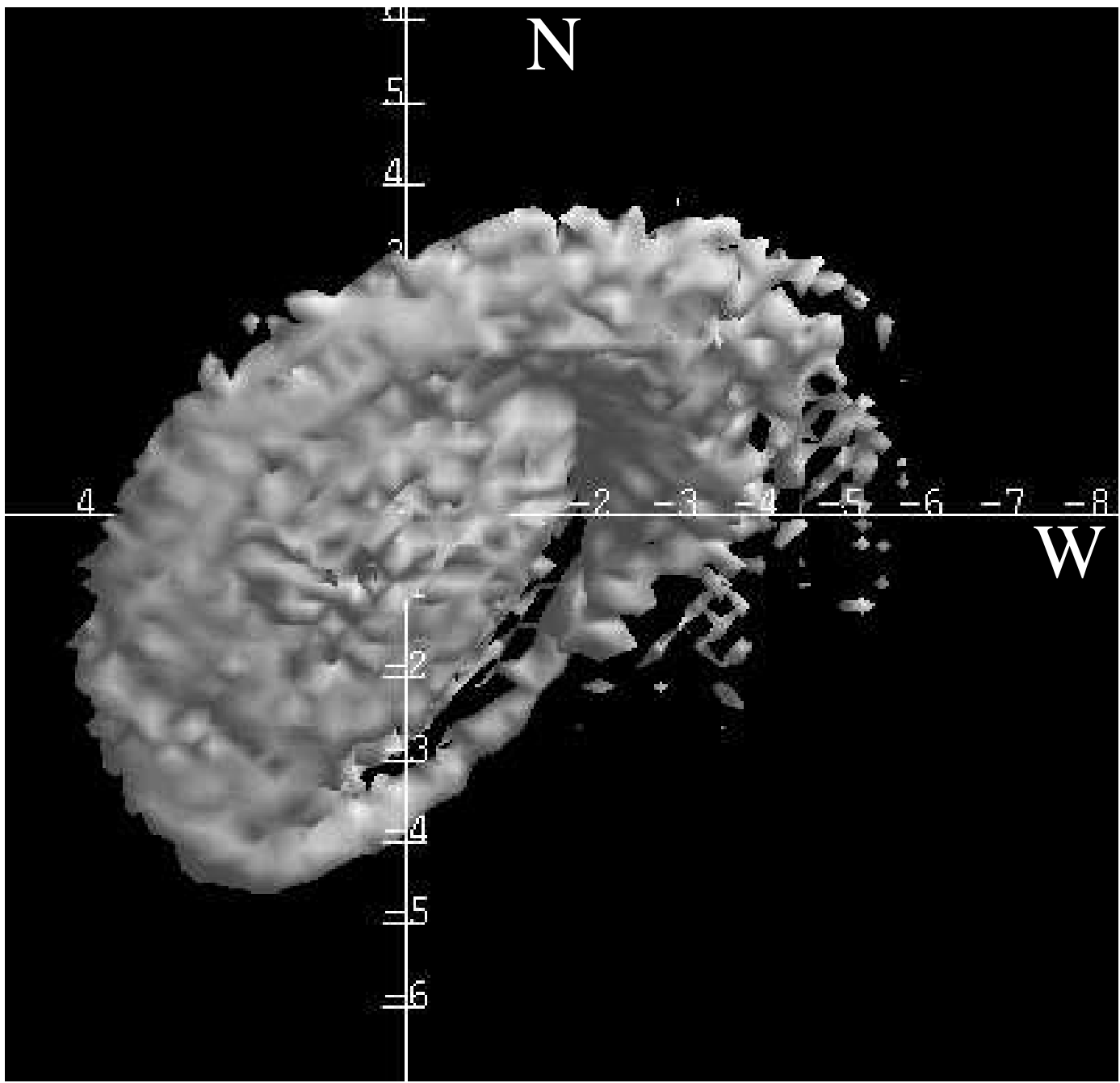
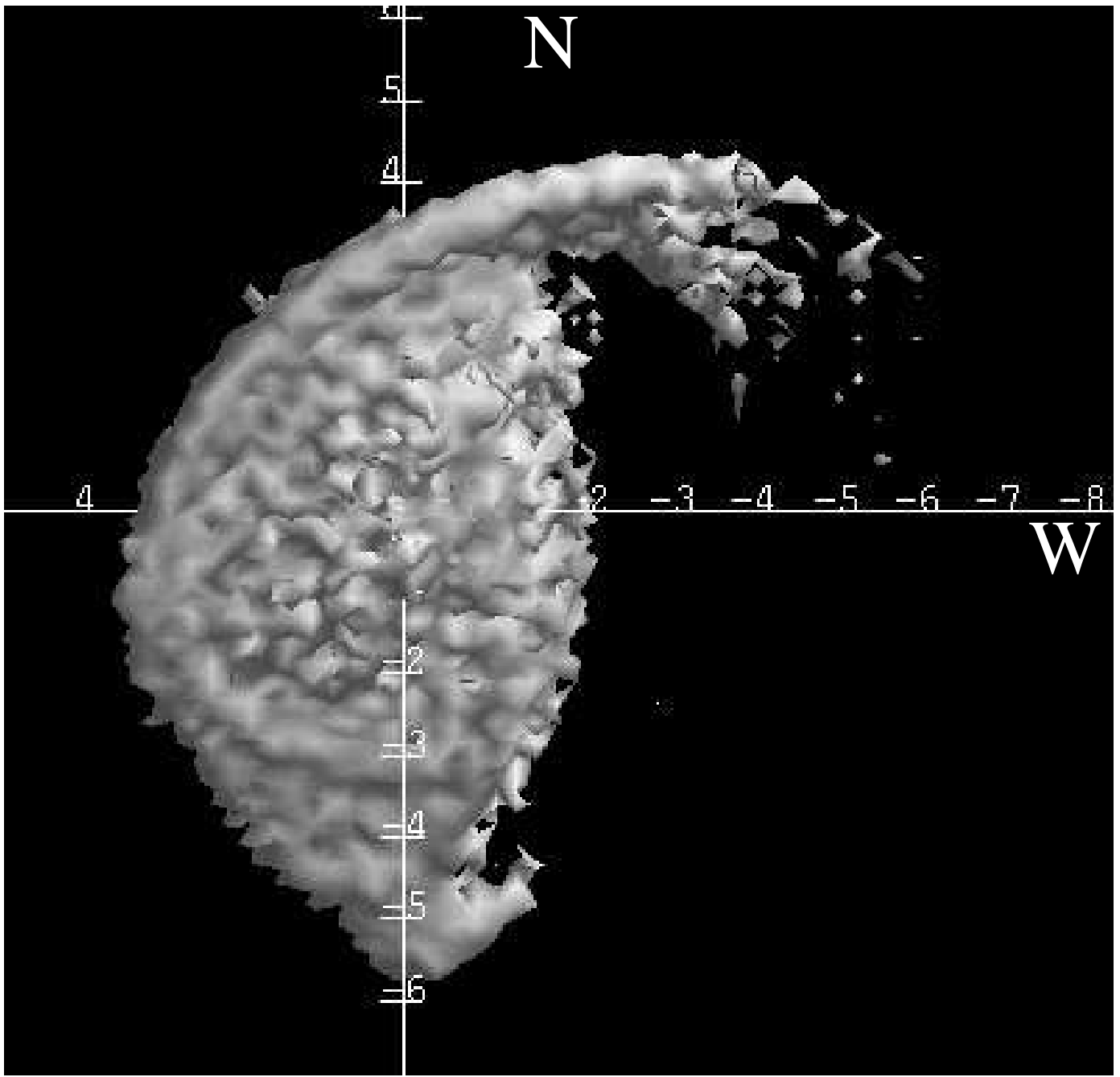
\end{document}